\documentclass[aps,prl,twocolumn,superscriptaddress,groupedaddress]{revtex4}  
\usepackage{graphicx}  
\usepackage{dcolumn}   
\usepackage{bm}        
\usepackage{amssymb}   
\usepackage[latin9]{inputenc}
\usepackage{amsmath}
\usepackage{subfigure}
\usepackage{amssymb}
\usepackage{graphicx}
\usepackage{amsfonts}%
\usepackage{dcolumn}
\usepackage{bm}
\usepackage{hyperref}
\usepackage{nccmath}
\usepackage{lipsum}
\newenvironment{mpmatrix}{\begin{medsize}\begin{pmatrix}}%
		{\end{pmatrix}\end{medsize}}%

\begin{document}
	\title{\bf Inducing 3-component fermions in centrosymmetric system by breaking TRS}
	\author{Chi-Ho Cheung}\email{f98245017@ntu.edu.tw}
	\affiliation{Department of Physics, National Taiwan University, Taipei 10617, Taiwan}
	\author{R. C. Xiao}\email{xiaoruichun@foxmail.com}
	\affiliation{Key Laboratory of Materials Physics, Institute of Solid State Physics, Chinese Academy of Sciences, Hefei 230031, China}
	\author{Ming-Chien Hsu}
	\affiliation{Department of physics, National Sun Yat-Sen University}
	\author{Huei-Ru Fuh}
	\affiliation{Department of Physics, National Taiwan University, Taipei 10617, Taiwan}
	\author{Yeu-Chung Lin}
	\affiliation{Department of Physics, National Taiwan University, Taipei 10617, Taiwan}
	\author{Ching-Ray Chang}
	\affiliation{Department of Physics, National Taiwan University, Taipei 10617, Taiwan}

	\date{\today}

	\begin{abstract}
    Recent researches show that by breaking inversion symmetry Dirac fermions can split into new fermions with 3-component. In this article, we demonstrate that Dirac fermions can also split into 3-component fermions with time reversal symmetry (TRS) breaking while inversion symmetry is preserved. Firstly, we conduct a symmetry analysis with the commutation relations among all symmetry operators of a Dirac semimetal and find out the symmetry conditions of Dirac fermions splitting into 3-component fermions. With the symmetry conditions, we derive the $k\cdot P$ effective Hamiltonian of TRS breaking and compare it with the Hamiltonian of inversion symmetry breaking. We find that they are different in $\Gamma$ point eigenenergies. This can be considered as consequence of Kramers degeneracy breaking which is a clear signature of TRS breaking. Moreover, with the $k\cdot P$ effective Hamiltonian of the system we show that TRS-breaking-induced 3-component fermions can split into Weyl fermions while a small magnetic field is applied. At the end, we show our first principles calculation results are consistent with the symmetry analysis and the $k\cdot P$ predictions. Our work clarifies the similarities and the differences between TRS-breaking-induced 3-component fermions and inversion-symmetry-breaking-induced 3-component fermions and extends the scope of 3-component fermion behaviors.

	\end{abstract}
	
	\maketitle
	
\begin{figure}[b]
			\centering
			\subfigure []{
				\includegraphics[width=3.5cm]{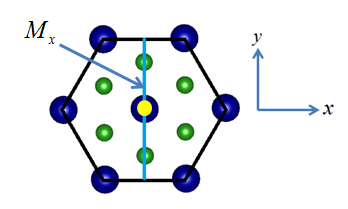}
				\label{Fig: 1a} }
			\subfigure []{
				\includegraphics[width=3.5cm]{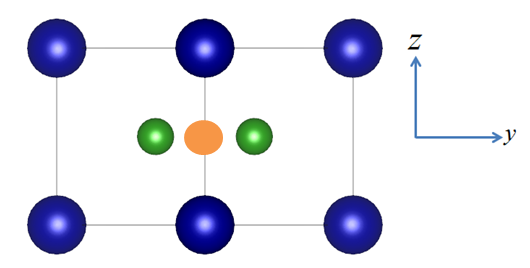}
				\label{Fig: 1b}}
			\subfigure []{
				\includegraphics[width=2.8cm]{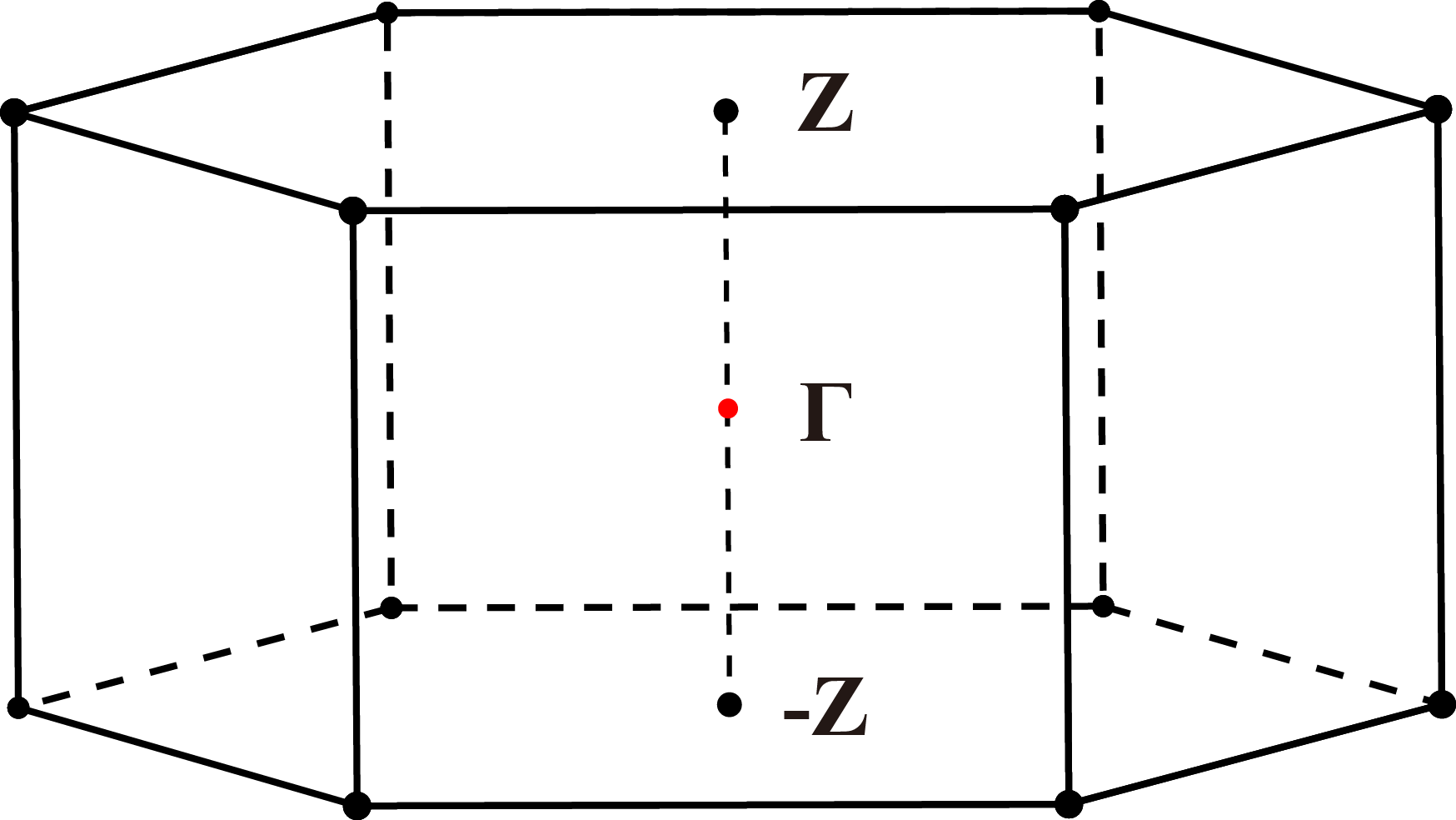}
				\label{Fig: 1c} }
			\caption{a) shows the top view of crystal structure with $D_{6h}$ point group symmetry. The $C_{3z}$ and $C_{2z}$ rotational symmetry operations rotate about the $z$ axis which is located at the center of the hexagon (marked by yellow circle). The mirror plane of the reflection symmetry-$M_{x}$ is marked in blue. b) is the profile of the crystal structure. The center point of inversion symmetry is on the $z$ axis (marked in orange circle). c) shows the first Brillouin zone of the primitive unit cell.
			}\label{Fig: 1}
		\end{figure}

	\emph{Symmetry analysis.---}It is well known that Dirac points can exist in a system which has both $D_{6h}$ point group symmetry and TRS\cite{Na3Bi}. $D_{6h}$ point group symmetry includes inversion, $C_{3z}$, $M_{x}$ and $C_{2z}$ symmetries (as shown in Fig. \ref{Fig: 1}). Since the system has inversion symmetry and TRS, all bands at any $k$ point have spin degeneracy. As Dirac points are a crossing point of two spin up-down degenerate bands, Dirac points have 4-fold degeneracy.

	It is necessary to break inversion symmetry or TRS to split the spin degeneracy and turn the Dirac points into 3-component fermions. Other references have shown the case of discovering 3-component fermions in non-centrosymmetric system with TRS\cite{1, 2, 3, I1, I2, I3, I4, I5, I6, I7, I8} and the case of inducing 3-component fermions by inversion symmetry breaking term while system does not have TRS\cite{4}. In this paper, we focus on discussing the case of preserving inversion and breaking TRS to induce 3-component fermions.
	
	The basis functions of Dirac points are $|S^{\pm}_{\frac{1}{2}},+\frac{1}{2}>$, $|S^{\pm}_{\frac{1}{2}},-\frac{1}{2}>$, $|P^{\pm}_{\frac{3}{2}},+\frac{3}{2}>$, $|P^{\pm}_{\frac{3}{2}},-\frac{3}{2}>$, the superscripts $\pm$ indicate the parity of wave function, $\pm1/2$ and $\pm3/2$ are labeled for $spin=\pm1/2$ and $spin=\pm3/2$ respectively. Note that the basis functions of Dirac points in Ref. \cite{Na3Bi} are $|S^{+}_{\frac{1}{2}},+\frac{1}{2}>$, $|S^{+}_{\frac{1}{2}},-\frac{1}{2}>$, $|P^{-}_{\frac{3}{2}},+\frac{3}{2}>$, $|P^{-}_{\frac{3}{2}},-\frac{3}{2}>$. However, the following discussion is generalizable to the $S$ orbital with negative parity and $P$ orbital with positive parity, thus we use superscript $\pm$ instead of superscript $+$ or $-$.
	
	With the basis functions, the matrix form of the symmetry operators of $D_{6h}$ point group and the matrix form of time reversal operator (TRO) are as follows:
	\begin{equation} \label{eq:1}
\begin{split}
\begin{array}{lcl}
	for \ basis:
	|S^{\pm}_{\frac{1}{2}},+\frac{1}{2}>, \ |S^{\pm}_{\frac{1}{2}},-\frac{1}{2}> \\
	   \ \ \ \ \ C_{3z} \ \ \ \ \ \ \ \ \ \  C_{2z}  \ \ \ \ \ \ \ \  M_{x}  \ \ \ \  Inversion \ \ \ \ \  TRO  \\
		{\begin{matrix}
		\begin{bmatrix}
		e^{\frac{i\pi}{3}} & 0 \\
		0 & e^{\frac{-i\pi}{3}}
		\end{bmatrix} &  \begin{bmatrix}
		i & 0 \\
		0 & -i
		\end{bmatrix} &  \pm\begin{bmatrix}
		0 & i \\
		i & 0
		\end{bmatrix} &  \pm\begin{bmatrix}
		1 & 0 \\
		0 & 1
		\end{bmatrix} &  \pm\begin{bmatrix}
		0 & -1 \\
		1 & 0
		\end{bmatrix}K
    	\end{matrix}}, \\
    \\
    for \ basis:
    |P^{\pm}_{\frac{3}{2}},+\frac{3}{2}>, \ |P^{\pm}_{\frac{3}{2}},-\frac{3}{2}> \\
    \ \ \ \ \  C_{3z} \ \ \ \ \ \ \ \ \  C_{2z}  \ \ \ \ \ \ \ \ \  M_{x}  \ \ \ \ \  Inversion \ \ \ \ \ \ \  TRO  \\
    {\begin{matrix}
    	\begin{bmatrix}
    	e^{i\pi} & 0 \\
    	0 & e^{-i\pi}
    	\end{bmatrix} &  \begin{bmatrix}
    	-i & 0 \\
    	0 & i
    	\end{bmatrix} &  \mp\begin{bmatrix}
    	0 & i \\
    	i & 0
    	\end{bmatrix} &  \mp\begin{bmatrix}
    	-1 & 0 \\
    	0 & -1
    	\end{bmatrix} &  \pm\begin{bmatrix}
   		0 & 1 \\
   		-1 & 0
   		\end{bmatrix}K
   		\end{matrix}},
\end{array}
\end{split}
    	\end{equation}
	where $K$ is complex conjugate operator.
	
	With TRS and $D_{6h}$ point group symmetry, Dirac points (crossing points) locate at the $\Gamma-Z$ axis\cite{Na3Bi}.
	
	We start the symmetry reduction process by breaking TRS first while keeping inversion symmetry intact.
	
	Any point group symmetry operator $S$ acts on a $k_{h}$ vector, such that the $Sk_{h}=k_{h}+nG$, where $G$ is any reciprocal lattice vector and $n$ is an integer number which is greater than or equal to zero, then those symmetry operators form the little group of the $k_{h}$ vector. Furthermore, Hamiltonian $H(k_{h})$ has to commute with all the symmetry operators of this little group. If any symmetry operator of this little group does not commute with each other in a subspace, then $H(k_{h})$ has to be degenerate in this subspace, otherwise $H(k_{h})$ cannot commute with all the symmetry operators of this little group simultaneously.
	
	Any $k$ point on $\Gamma-Z$ axis-$k_{z}$ is invariant under $C_{3z}$, $C_{2z}$ rotation or $M_{x}$ reflection, thus $\Gamma-Z$ axis has $C_{6v}$ point group symmetry. Furthermore, $C_{3z}$ and $C_{2z}$ do not commute with $M_{x}$ in $spin=\pm1/2$ subspace, $C_{2z}$ and $M_{x}$ do not commute with each other in $spin=\pm3/2$ subspace (as shown in Eq. \ref{eq:1}). Therefore $spin=\pm1/2$ subspace and $spin=\pm3/2$ subspace still remain two fold degenerate after TRS breaking. Since $C_{3z}$ commute with $C_{2z}$ in $spin=\pm1/2$ subspace and $spin=\pm3/2$ subspace, breaking $M_{x}$ will split $spin=\pm1/2$ subspace and $spin=\pm3/2$ subspace. For inducing 3-component fermions, one of these two degeneracies should be remained while the other one splits off. Thus breaking $M_{x}$ is not an option. Besides, $C_{3z}$ symmetry can protect the crossing point on $\Gamma-Z$ axis\cite{Na3Bi}, and it is undesirable to move the crossing point away from $\Gamma-Z$ axis or gapping it, thus $C_{3z}$ symmetry are chosen to be preserved.

	Base on the symmetry analysis above, breaking $C_{2z}$ is the only option left. When $C_{2z}$ symmetry and TRS are broken, $D_{6h}+TRS$ is reduced to $D_{3d}$ and then all the symmetry operators of the little group of $k_{z}$ vector in $spin=\pm3/2$ subspace commute with each other. Therefore $spin=\pm3/2$ shall be split off on the $\Gamma-Z$ axis. On the other hand, $C_{3z}$ does not commute with $M_{x}$ in $spin=\pm1/2$ subspace. Thus the line degeneracy of $spin=\pm1/2$ subspace on $\Gamma-Z$ axis shall remain intact. Hence, we can expect that Dirac fermions split into 3-component fermions when $C_{2z}$ symmetry and TRS are broken.
	
	Actually, combining $C_{2z}$ symmetry and TRS to one symmetry, $C_{2z}\cdot TRO$ symmetry, can achieve the goal as well, because $C_{2z}\cdot TRO$ and $M_{x}$ have the same set of eigenvectors and $spin=\pm3/2$ subspace is reducible. In such case, symmetry of the system is reduced from $D_{6h}+TRS$ to $D_{3d}+(C_{2z}\cdot TRO)$. Note that after the symmetry combination, $C_{2z}$ symmetry alone or TRS alone is no longer a symmetry of the system. Thus the symmetry combination breaks $C_{2z}$ and TRS as well. However the symmetry group $D_{3d}+(C_{2z}\cdot TRO)$ is larger than $D_{3d}$.

\begin{widetext}
		\begin{figure*}
			\centering
			\includegraphics[width=15.0cm,height=5.0cm]{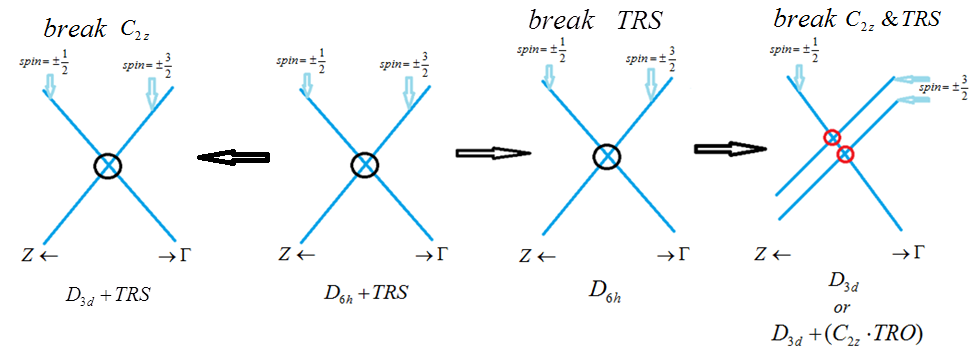}
			\caption{The schematic figure of the symmetry reduction processes of the system. When $D_{6h}+TRS$ reduce to $D_{6h}$ or $D_{3d}+TRS$, Dirac point (marked in black circle) does not split. When the symmetry of the system becomes $D_{3d}$ or $D_{3d}+(C_{2z}\cdot TRO)$, Dirac point split into 3-component fermions (marked in red circles).
			}\label{Fig: 2}
		\end{figure*}
	\end{widetext}

	The schematic figure of the symmetry reduction processes is shown in Fig. \ref{Fig: 2}.
	
	\emph{$k\cdot P$ perturbation theory.---}
	In this section, we calculate $k\cdot P$ effective Hamiltonian to show how the Dirac points split into 3-component fermions while $C_{2z}$ symmetry and TRS are broken. On the other hand, we compare the TRS breaking Hamiltonian with the inversion symmetry breaking Hamiltonian in order to see the differences between them. Moreover, we prove that TRS-breaking-induced 3-component fermions can split into Weyl points while applying a small uniform magnetic field $B_{z}$ in $z$ direction.
	
	The symmetry conditions-$SH(k')S^{-1}=H(S^{-1}k')$ are used to build the $k\cdot P$ effective Hamiltonian, where $S$ is any symmetry operator of the system symmetry, $k'=\alpha k_{z}^{l}k_{+}^{m}k_{-}^{n}$, $\alpha$ is a $k$ independent coefficient, $k_{\pm}=k_{x}\pm ik_{y}$, and $l$, $m$, $n$ are integers which are greater than or equal to zero. The coefficient $\alpha$ of any $k'$ term can be non-zero when the $k'$ term matches the symmetry condition, otherwise $\alpha$ has to be zero.
	
	In this article, the order of the basis functions of $k\cdot P$ effective Hamiltonian are chosen to be $|S^{\pm}_{\frac{1}{2}},+\frac{1}{2}>$, $|S^{\pm}_{\frac{1}{2}},-\frac{1}{2}>$, $|P^{\pm}_{\frac{3}{2}},+\frac{3}{2}>$, $|P^{\pm}_{\frac{3}{2}},-\frac{3}{2}>$.

	When the $C_{3z}$ operator acts on the effective Hamiltonian $C_{3z}H_{C_{3}}(k)C^{-1}_{3z}$, according the $C_{3z}$ symmetry operator in Eq. \ref{eq:1}, a phase factor $e^{\frac{2i\pi}{3}}$ or $-e^{\frac{i\pi}{3}}$ will be generated on the matrix elements of off-diagonal block and the matrix elements $H_{12}$, $H_{21}$. Therefore these matrix elements must have $k_{+}$ or $k_{-}$ factor to match the $C_{3z}$ symmetry condition $C_{3z}H_{C_{3}}(k')C^{-1}_{3z}=H_{C_{3}}(C^{-1}_{3z}k')$. If we only consider the dispersion relations on $\Gamma-Z$ axis, all these matrix elements become zero.
	
	After $C_{2z}$ symmetry and TRS are broken, $D_{6h}+TRS$ becomes $D_{3d}$. The effective Hamiltonian with $D_{3d}$ point group symmetry on $\Gamma-Z$ axis is given as follows:
		\begin{equation} \label{eq:2}
		\begin{array}{lcl}
		H_{D_{3d}}(k_{z})=\varepsilon(k_{z})+ \\
		\begin{mpmatrix}
		M_{0}-M_{1}k^{2}_{z} & 0 & 0 & 0 \\
		0 & M_{0}-M_{1}k^{2}_{z} & 0 & 0 \\
		0 & 0 & -M_{0}+M_{1}k^{2}_{z} & C \\
		0 & 0 & C & -M_{0}+M_{1}k^{2}_{z} \\
		\end{mpmatrix},
		\end{array}
		\end{equation}
	where the expansion is only up to the first order of $k$ for off-diagonal matrix elements and up to the second order of $k$ for diagonal matrix elements. $\varepsilon(k_{z})=D_{0}+D_{1}k_{z}^{2}$. $M_{0}$ and $M_{1}$ are real positive $k$ independent coefficients. $D_{0}$, $D_{1}$ and $C$ are real $k$ independent coefficients. Comparing the Hamiltonian in Eq. \ref{eq:2} with the Hamiltonian of $D_{6h}+TRS$ in Ref. \cite{Na3Bi} (note that the basis functions herein are not in same order as presented in Ref. \cite{Na3Bi}), we find that simultaneously breaking of $C_{2z}$ symmetry and TRS induce $C(\tau_{0}\sigma_{x}-\tau_{z}\sigma_{x})/2$ ($\tau$ is in orbital space, $\sigma$ is in spin space). As $C(\tau_{0}\sigma_{x}-\tau_{z}\sigma_{x})/2$ can also survive under the symmetry condition of $(C_{2z}\cdot TRO)H(k')(C_{2z}\cdot TRO)^{-1}=H((C_{2z}\cdot TRO)^{-1}k')$, therefore $H_{D_{3d}}(k_{z})$ in Eq. \ref{eq:2} is equal to $H_{D_{3d}+(C_{2z}\cdot TRO)}(k_{z})$.
	
	
	The eigenvalues of $H_{D_{3d}}(k_{z})$ are:
	\begin{equation} \label{eq:eigen1}
	\begin{array}{lcl}
	E_{1}=E_{2}=\varepsilon(k_{z})+M_{0}-M_{1}k^{2}_{z},  \\
	\\
	E_{3}, E_{4}=\varepsilon(k_{z})-M_{0}+M_{1}k^{2}_{z}\pm C,
	\end{array}
	\end{equation}
	where the symmetry breaking term-$C$ breaks the line degeneracy of $spin=\pm3/2$ subspace just as expected. With conditions of $E_{1},$ $E_{2}=E_{3}$ and $E_{1},$ $E_{2}=E_{4}$, we get the crossing point positions-$k_{C_{1}},$ $k_{C_{2}},$ $k_{C_{3}},$ $k_{C_{4}}=\pm\sqrt{\frac{2M_{0}\pm C}{2M_{1}}}$, which indicates that the two Dirac points split into four 3-component fermions.
	
\begin{figure}[t]
		\centering
		\subfigure []{
			\includegraphics[width=4.0cm]{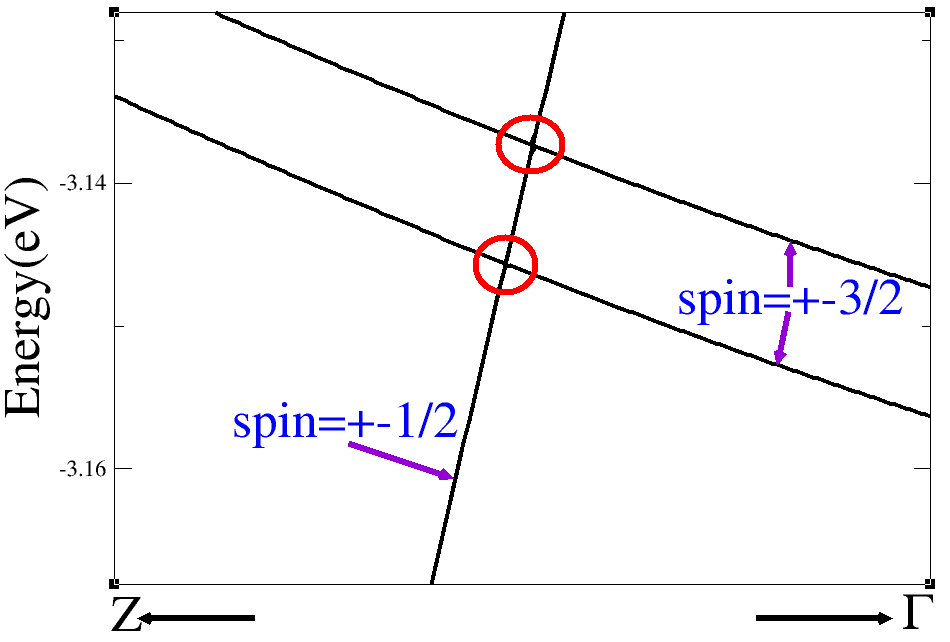}
			\label{Fig: 4a} }
		\subfigure []{
			\includegraphics[width=4.0cm]{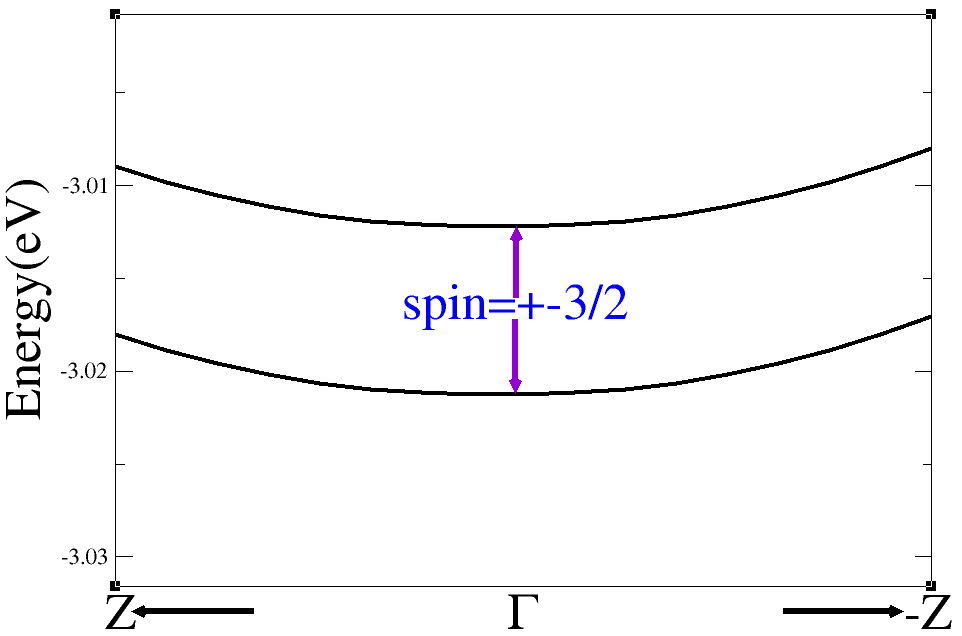}
			\label{Fig: 4b}}
		\subfigure []{
			\includegraphics[width=4.0cm]{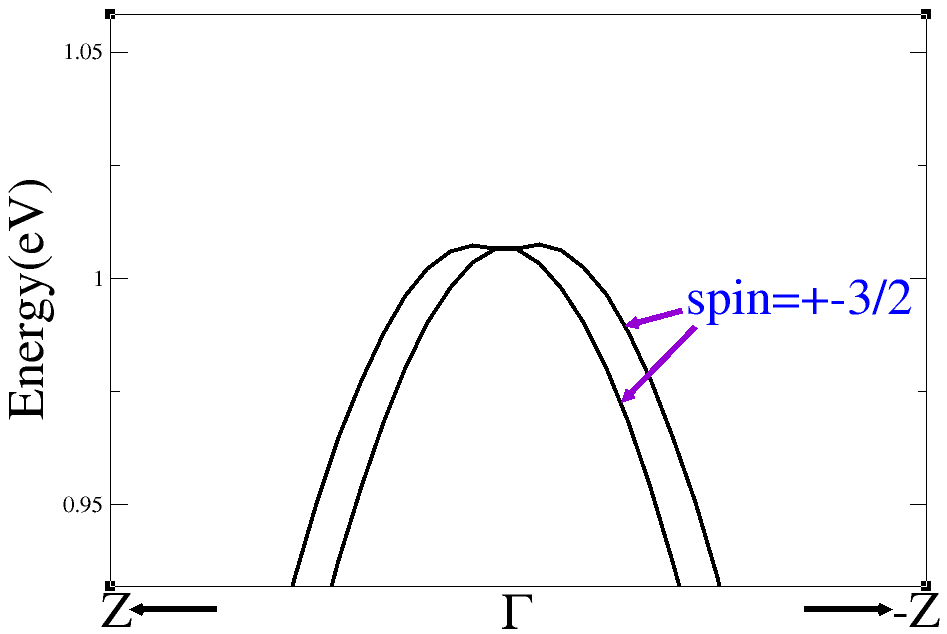}
			\label{Fig: 4c} }
		\caption{The band structures from first principles calculation. a) shows that 3-component fermions (marked in red circles) can exist in a system which has $6'/m'mm'$ magnetic point group symmetry due to the band crossing between doublet $spin=\pm1/2$ and singlet $spin=\pm3/2$. b) shows that $spin=\pm3/2$ subspace split on $\Gamma-Z$ axis (including $\Gamma$ point) while the system has $6'/m'mm'$ magnetic point group symmetry. c) shows that $spin=\pm3/2$ subspace splits on $\Gamma-Z$ axis (but degenerates at $\Gamma$ point) in $MoP$.
		}\label{Fig: 4}
	\end{figure}

	For the case in $MoP$\cite{1, 3}, 3-component fermions are induced by inversion symmetry breaking. $MoP$ has $D_{3h}$ point group symmetry and TRS. Its effective Hamiltonian on $\Gamma-Z$ axis is given as follows:
	\begin{equation} \label{eq:4}
	\begin{array}{lcl}
	H_{D_{3h}}(k_{z})=\varepsilon(k_{z})+ \\
	\begin{mpmatrix}
	M_{0}-M_{1}k^{2}_{z} & 0 & 0 & 0 \\
	0 & M_{0}-M_{1}k^{2}_{z} & 0 & 0 \\
	0 & 0 & -M_{0}+M_{1}k^{2}_{z} & Bk_{z} \\
	0 & 0 & Bk_{z} & -M_{0}+M_{1}k^{2}_{z} \\
	\end{mpmatrix},
	\end{array}
	\end{equation}
	where $B$ is a real $k$ independent coefficient and the expansion is only up to the first order of $k$ for off-diagonal matrix elements and up to the second order of $k$ for diagonal matrix elements. And the eigenvalues of $H_{D_{3h}}(k_{z})$ are:
	\begin{equation} \label{eq:eigen2}
	\begin{array}{lcl}
	E_{1}=E_{2}=\varepsilon(k_{z})+M_{0}-M_{1}k^{2}_{z},  \\
	\\
	E_{3}, E_{4}=\varepsilon(k_{z})-M_{0}+M_{1}k^{2}_{z}\pm Bk_{z}.
	\end{array}
	\end{equation}
	The dispersion relationships in Eq. \ref{eq:eigen2} is consistent with our first principles calculation of $MoP$ (as shown in Fig. \ref{Fig: 4}). Comparing Eq. \ref{eq:eigen1} with Eq. \ref{eq:eigen2}, we find that TRS breaking splits the line degeneracy of $spin=\pm3/2$ including $\Gamma$ point, while inversion symmetry breaking splits the line degeneracy of $spin=\pm3/2$ without $\Gamma$ point splitting. Such a difference in $\Gamma$ point eigenenergies can be considered as consequence of Kramers degeneracy breaking in the system without TRS while $MoP$ has TRS. The crossing point positions of inversion symmetry breaking are $k_{C_{1}},$ $k_{C_{2}},$ $k_{C_{3}},$ $k_{C_{4}}=\frac{\pm\frac{B}{2}\pm\sqrt{\frac{B^{2}}{4}+4M_{1}M_{0}}}{-2M_{1}}$.
	
	Back to the case of TRS breaking, if a small uniform magnetic field $B_{z}$ is applied in $z$ direction, $D_{3d}$ symmetry or $D_{3d}+(C_{2z}\cdot TRO)$ symmetry will be reduced to $C_{3i}$ symmetry. Since $M_{x}$ and $(C_{2z}\cdot TRO)$ are broken by the magnetic field, $spin=\pm1/2$ subspace shall split. Thus the 3-component fermions shall split into Weyl points. From the point of view of $k\cdot P$ effective Hamiltonian, $B_{z}$ induces $k$ independent terms $\Delta_{1}$, $\Delta_{2}$, $\Delta_{3}$, $\Delta_{4}$, $\Delta_{5}$ and $\Delta_{5}^{*}$ in matrix elements $H_{11}$, $H_{22}$, $H_{33}$, $H_{44}$, $H_{34}$ and $H_{43}$ respectively (because $B_{z}$ is small, only zeroth order terms are considered). Adding these induced terms to the original effective Hamiltonian $H_{D_{3d}}(k_{z})$, we can solve the new eigenenergies, and then get the new crossing point positions $k_{C_{1}},$ $k_{C_{2}},$ $k_{C_{3}},$ $k_{C_{4}}=\pm\sqrt{\frac{2M_{0}+(\Delta_{1}-\Delta _{A} )\pm \Delta_{C}}{2M_{1}}},$ $k_{C_{5}},$ $k_{C_{6}},$ $k_{C_{7}},$ $k_{C_{8}}=\pm\sqrt{\frac{2M_{0}+(\Delta_{2}-\Delta _{A} )\pm \Delta_{C}}{2M_{1}}}$, where $\Delta_{A}=\frac{\Delta_{3}+\Delta_{4}}{2}$, $\Delta_{C}=\sqrt{(\frac{\Delta_{3}-\Delta_{4}}{2})^{2}+[Re(\Delta_{5})+C]^{2}+[-Im(\Delta_{5})]^{2}}$, $Re(\Delta_{5})$ is the real part of $\Delta_{5}$ and $Im(\Delta_{5})$ is the imaginary part of $\Delta_{5}$. After applying $B_{z}$, 4 crossing points become 8 crossing points. Since $spin=\pm1/2$ subspace and $spin=\pm3/2$ subspace are non-degenerate, each crossing point is 2-fold degenerate. This is the evidence that 3-component fermions split into Weyl points. Furthermore, using $k\cdot P$ expansion around the crossing points can prove that the 8 crossing points have 4 pairs of Weyl points with opposite chiralities (as shown in Supplementary materials). Thus we show that TRS-breaking-induced 3-component fermions can split into Weyl fermions while applying a small magnetic field in $z$ direction. This splitting behavior is just like that of inversion-symmetry-breaking-induced 3-component fermions\cite{3}.

	

	\emph{First principles calculation.---}
	In this section, firstly we use first principles calculation result to prove that a system with magnetic point group symmetry of $6'/m'mm'$ can have 3-component fermions and inversion symmetry simultaneously. Secondly, we compare the band structure of the system with the band structure of $MoP$ ($MoP$ has TRS but no inversion symmetry).

	\begin{figure}[b]
		\centering
		\subfigure []{
			\includegraphics[width=3.5cm]{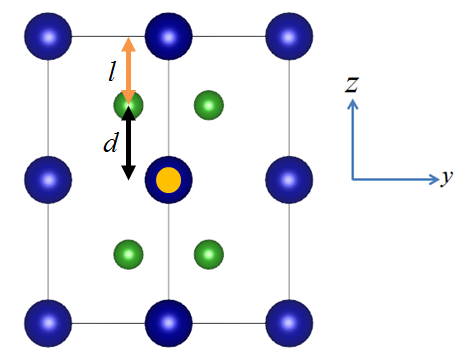}
			\label{Fig: 3a} }
		\subfigure []{
			\includegraphics[width=3.5cm]{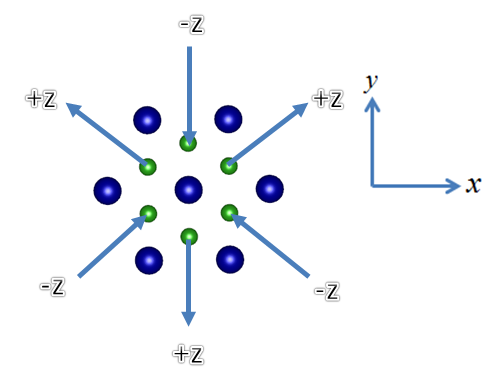}
			\label{Fig: 3b}}
		\subfigure []{
			\includegraphics[width=3.5cm]{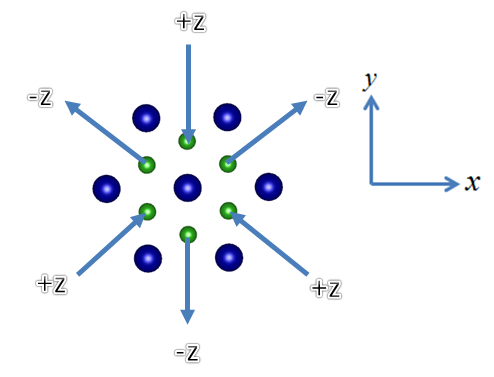}
			\label{Fig: 3c} }
		\caption{The symmetry of the system is reduced from $D_{6h}+TRS$ to $D_{3d}+(C_{2z}\cdot TRO)$. a) shows that due to the symmetry reduction, the unit cell in Fig. \ref{Fig: 1} is enlarged in $z$ direction, and the inversion center is shifted in $z$ direction (the new position of inversion center is marked with orange circle). The $x$, $y$ components of the magnetic moments at b) upper layer and at c) lower layer are marked with blue arrows. The $z$ component of the magnetic moments are marked by $+z$ or $-z$. We assume that all magnetic moments are on green atoms and violet blue atoms do not have any magnetic moment.
		}\label{Fig: 3}
	\end{figure}

	$6'/m'mm'$ is one of the magnetic point group of $D_{6h}$ point group. It has inversion symmetry, $M_{x}$ symmetry, $C_{3z}$ symmetry, but the $C_{2z}$ symmetry is combined with TRS.
	
	Since all the ferromagnetic always generate a magnetic field that breaks either $C_{3z}$ symmetry or $M_{x}$ symmetry, thus $6'/m'mm'$ has to be antiferromagnetic. Its crystal structure and magnetic moments are shown in Fig. \ref{Fig: 3}. Note that the length of $l$ and the length of $d$ (shown in Fig. \ref{Fig: 3a}) must be different if the magnetic moments do not have $x$, $y$ components, otherwise the system will have $\left \{C_{2z}|0 \ 0 \ 1/2\right \}$ ($C_{2z}$ rotation followed by a translation of half an unit cell vector in $z$ direction) symmetry which protects the line degeneracy of $spin=\pm3/2$ subspace on $\Gamma-Z$ axis and forbids Dirac points to split into 3-component fermions.
	
	All first principles calculations are performed by using Vienna Ab initio Simulation Package (VASP) code\cite{VASP1, VASP2}. The generalized gradient approximation (GGA) and the Perdew-Burke-Ernzerhof (PBE) exchange-correlation functional along with projector augmented wave method are used for self-consistent total energy calculations and band calculations. A plane-wave basis set with an energy cutoff of 500 eV is employed. The energy convergence criteria for electronic and ionic iterations are set to be $10^{-4}$ eV. Using the Gamma-centered grid method, the reciprocal space is meshed at $8 \times 8 \times 4$ for $6'/m'mm'$ magnetic point group system ($8 \times 8 \times 8$ for $MoP$). Binary compound-chromium borides ($CrB_{2}$) is used for the simulations of $6'/m'mm'$ magnetic point group system, directions of spins are pinned at appropriate orientations to match the magnetic point group symmetry of $6'/m'mm'$ in all steps of calculations. In both of the cases ($CrB_{2}$ and $MoP$), $s$, $p$, $d$ orbitals and spin-orbit coupling are considered in self-consistent total energy calculations and band calculations.

	The band structures from first principles calculation indicate that 3-component fermions can exist in a system which has $6'/m'mm'$ magnetic point group symmetry due to the band crossing between doublet $spin=\pm1/2$ and singlet $spin=\pm3/2$ (as shown in Fig. \ref{Fig: 4a}). Besides, the $k\cdot P$ predictions for $\Gamma$ point behaviors of $spin=\pm3/2$ subspace in $MoP$ and in $6'/m'mm'$ system are well matched with first principles calculation results (as shown in Fig. \ref{Fig: 4b} and Fig. \ref{Fig: 4c}).

	\emph{Conclusion.---}It was well known that Dirac fermions can split into Weyl fermions by either breaking inversion symmetry or breaking TRS. Other recent researches show that by breaking inversion symmetry Dirac fermions may split into 3-component fermions instead of Weyl fermions. We show that 3-component fermions can also be induced by breaking TRS, which means 3-component fermions can be induced by inversion symmetry breaking or TRS breaking just like Weyl fermions.
	
	We conduct the symmetry analysis with the commutation relations between all the symmetry operators of the system and predict the existence of 3-component fermions before going into detail calculation.
	
	In $k\cdot P$ section, with the Hamiltonian of TRS breaking, we confirm that 3-component fermions can be induced by TRS breaking while inversion symmetry is held. Besides, comparing the Hamiltonian of TRS breaking with that of inversion symmetry breaking, we find that TRS breaking induces $\Gamma$ point splitting in $spin=\pm3/2$ subspace while inversion breaking does not. Such a different splitting behavior in $\Gamma$ point is a direct consequence of Kramers degeneracy breaking in the system with TRS breaking. Moreover, we show that TRS-breaking-induced 3-component fermions can split into Weyl fermions while a small magnetic field in $z$ direction is applied. This splitting behavior is just like that of inversion-symmetry-breaking-induced 3-component fermions. Hence, our work clarifies the similarities and the differences between TRS-breaking-induced 3-component fermions and inversion-symmetry-breaking-induced 3-component fermions and fills the blank of 3-component fermion behaviors.
	
	In first principles calculation section, we show how to simulate the case of TRS breaking by first principles calculation. This is useful for further theoretical study, since it suggests one of the ways to simulate "pure" TRS-breaking-induced 3-component fermions numerically before any real material is found. In addition, the first principles calculation results support the symmetry analysis and the $k\cdot P$ predictions.

	\begin{center}
		\textbf{\large ACKNOWLEDGMENTS}
	\end{center}
	
	This work was studied supported by the Ministry of Science and Technology of Taiwan under Grant No. MOST 104-2112-M-002-007-MY3. We are grateful to Computer and Information Networking Center, National Taiwan University for the support of high-performance computing facilities.
	
	Chi-Ho Cheung and R. C. Xiao contributed equally to this work.

\end{document}